# Electronic transport properties and hydrostatic pressure effect of FeSe$_{0.67}$Te$_{0.33}$ single crystals free of phase separation

Xiangzhuo Xing,[1,6] Yue Sun,[1,2,6] Xiaolei Yi,[1] Meng Li,[1] Jiajia Feng,[1] Yan Meng,[1] Yufeng Zhang,[1] Wenchong Li,[1] Nan Zhou,[1] Xiude He,[3] Jun-Yi Ge,[3] Wei Zhou,[4] Tsuyoshi Tamegai,[5] and Zhixiang Shi[1,6]

[1]School of Physics, Southeast University, Nanjing 211189, China

[2]Department of Physics and Mathematics, Aoyama Gakuin University, Sagamihara 252-5258, Japan

[3]Materials Genome Institute, Shanghai University, Shanghai 200444, China

[4]Advanced Functional Materials Laboratory and Department of Physics, Changshu Institute of Technology, Changshu 215500, China

[5]Department of Applied Physics, The University of Tokyo, Tokyo 113-8656, Japan

[6]Authors to whom any correspondence should be addressed

E-mail: xzxing@seu.edu.cn (X. Xing), sunyue@phys.aoyama.ac.jp (Y. Sun), and zxshi@seu.edu.cn (Z. Shi)

## Abstract

FeSe$_{1-x}$Te$_x$ superconductors manifest some intriguing electronic properties depending on the value of $x$. In FeSe single crystal, the nematic phase and Dirac band structure have been observed, while topological surface superconductivity with the Majorana bound state was found in the crystal of $x \sim 0.55$. Therefore, the electronic properties of single crystals with $0 < x \leq 0.5$ are crucial for probing the evolution of those intriguing properties as well as their relations. However, this study is still left blank due to the lack of single crystals because of phase separation. Here, we report the synthesis, magnetization, electronic transport properties, and hydrostatic pressure effect of FeSe$_{0.67}$Te$_{0.33}$ single crystals free of phase separation. A structural (nematic) transition is visible at $T_s$ = 39 K, below which the resistivity exhibits a Fermi-liquid behavior. Analysis of the upper critical fields suggests that spin-paramagnetic effect should be taken into account for both $H \parallel c$ axis and $H \parallel ab$ plane. A crossover from the low-$H$ quadratic to the high-$H$ quasi-linear behavior is observed in the magnetoresistance, signifying the possible existence of Dirac-cone state. Besides, the strong temperature dependence of Hall coefficient, violation of (modified) Kohler's rule, and two-band model analysis indicate the multiband effects in FeSe$_{0.67}$Te$_{0.33}$ single crystals. Hydrostatic pressure measurements reveal that $T_s$ is quickly suppressed with pressure while $T_c$ is monotonically increased up to 2.31 GPa, indicating the competition between nematicity and superconductivity. No signature of magnetic order that has been detected in FeSe$_{1-x}$S$_x$ is observed. Our findings fill up the blank of the knowledge on the basic properties of FeSe$_{1-x}$Te$_x$ system with low-Te concentrations.

# 1. Introduction

The iron chalcogenide superconductor FeSe [1], which exhibits the simplest crystal structure among iron-based superconductors (IBSs), has attracted considerable interest due to its unique properties [2]. FeSe is a compensated semimetal that shows superconductivity (SC) with a superconducting transition temperature $T_c$ ~ 9 K at ambient pressure [1]. Intriguingly, $T_c$ can be significantly increased to 37 K under pressure [3]. More surprisingly, signs of SC with a $T_c$ exceeding 65 K have been observed in a monolayer FeSe thin film on the $SrTiO_3$ substrate [4, 5]. Because the Fermi energies are extremely small and comparable to the superconducting gap value, the SC in bulk FeSe has been argued to be close to a Bardeen–Cooper–Schrieffer/Bose–Einstein-condensation crossover [6]. Furthermore, unlike other IBSs, FeSe undergoes a structural (nematic) transition at $T_s$ ~ 87 K without any long-range magnetic order at ambient pressure [7-9]. Nevertheless, static magnetic order (spin density wave, SDW) can be stabilized under pressure [2, 10-14]. Hence, FeSe provides a fascinating platform for investigating the relationships among SC, nematicity, and magnetism.

Isovalent substitution, which maintains the nature of compensated semimetals of FeSe, is an ideal and effective route for tuning the ground state and electronic properties. Much of the recent attention has focused on the isovalent S doping that is equivalent to internal positive chemical pressure [2, 15]. The nematic order is gradually suppressed by S doping in $FeSe_{1-x}S_x$ and a non-magnetic nematic quantum critical point (QCP) appears at $x$ ~ 0.17 [2]. The normal state resistivity in the nematic phase exhibits $T$-linear behavior, indicating the non-Fermi-liquid behavior [16, 17]. Across the nematic QCP, the superconducting gap exhibits an abrupt change [18, 19] and quantum oscillation measurements have indicated a topological Lifshitz transition and a reduction in electronic correlations [20]. Very recently, we have constructed a complete doping phase diagram of $FeSe_{1-x}S_x$, which uncovers a characteristic temperature $T^*$ that shows a dome-shaped behavior and competes with SC in the high-S doping region [21].

In contrast to S substitution, another isovalent substitution of Te applies a negative chemical pressure due to its larger ionic radius. $T_c$ exhibits a maximum value of ~ 15 K in bulk $FeSe_{1-x}Te_x$ [22-24]. A bicollinear antiferromagnetic (AFM) order occurs in the high-Te doping region close to FeTe [22]. Unlike $FeSe_{1-x}S_x$ with highly anisotropic superconducting gaps [2, 25], a rather isotropic superconducting gap has been detected in $FeSe_{1-x}Te_x$ [26-28]. Intriguingly, recent studies have successfully observed the topological surface states [2, 27, 29] and Majorana bound states [2, 30, 31], which makes $FeSe_{1-x}Te_x$ the first high-temperature topological superconductor. However, thus far, experimental researches have been mainly performed in the high-Te doping region owing to the availability of $FeSe_{1-x}Te_x$ single crystals with $x > 0.5$ [32]. Synthesis of $FeSe_{1-x}Te_x$ single crystals in the low-Te doping region has been challenging due to the phase separation [22, 33]. Although the suppression of phase separation can be achieved in thin films using the pulsed laser deposition [34, 35], the presence of strain effect impedes the study of its intrinsic properties. Recently, Terao *et al.* reported the successful synthesis of $FeSe_{1-x}Te_x$ ($0 \leq x \leq 0.4$) single crystals by a flux method [36], which addresses the lack of single crystals in the phase separation region. Based on previous reports

[22, 36-38], a summarized phase diagram of FeSe$_{1-x}$Te$_x$ single crystals is depicted in figure 1. As can be seen, $T_s$ is linearly suppressed with Te doping and disappears at $x \sim 0.5$. Within the nematic phase, $T_c$ shows a nonmonotonic doping dependence with a minimum value at $x \sim 0.2$. The available single-phase crystals in the crossover region of $0 < x \leq 0.5$ allow us to probe the evolution of the above mentioned intriguing properties such as the nematic phase and topological surface state, and their relation (coexistence or mutual exclusivity; competing or supporting). Therefore, probing the basic properties of crystals in this region is of great importance and timely, which has not been performed well due to the difficulty in crystal growth.

In this work, we report the successful synthesis of FeSe$_{0.67}$Te$_{0.33}$ single crystals without phase separation by a flux method, which allows access to the intrinsic physical properties. The location of FeSe$_{0.67}$Te$_{0.33}$ in the phase diagram is represented by stars in figure 1. Comprehensive study of the resistivity, magnetization, upper critical field, Hall effect, magnetoresistance, and hydrostatic pressure measurements were performed. The obtained results reveal intrinsic properties of bulk FeSe$_{1-x}$Te$_x$ in the region of $0 < x \leq 0.5$ that have never been explored previously.

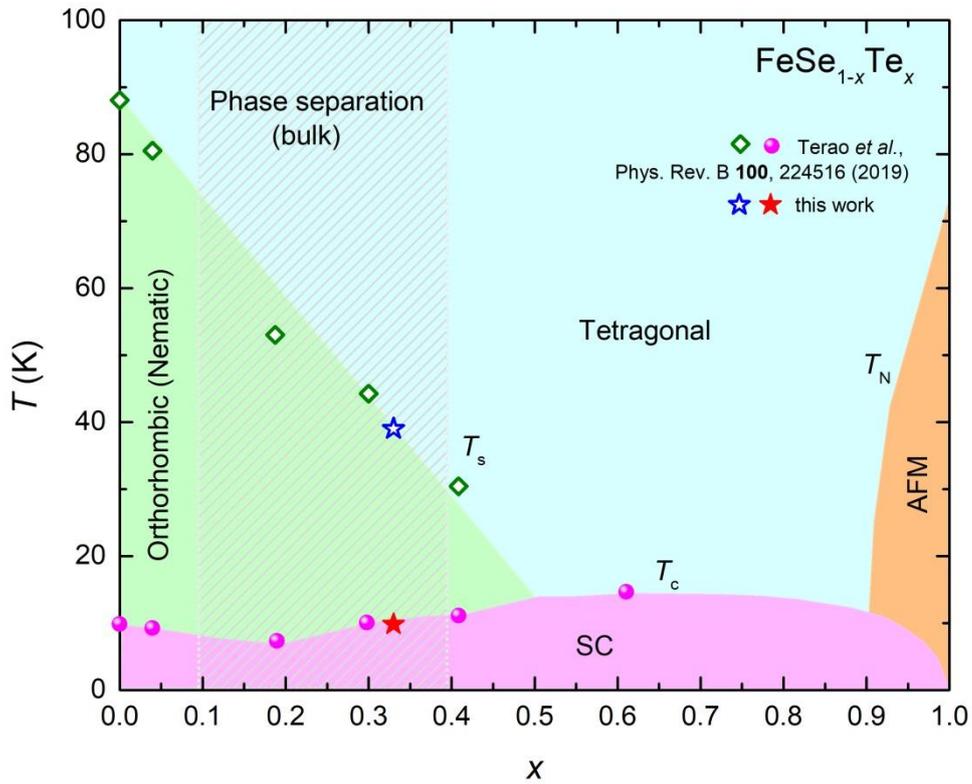

**Figure 1.** Complete *T-x* phase diagram of bulk FeSe$_{1-x}$Te$_x$, depicted based on previous reports [22, 36-38]. $T_s$, $T_c$, and $T_N$ denote the characteristic temperatures at which structural (nematic), superconducting, and AFM transitions occur, respectively. The red and blue stars represent the $T_c$ and $T_s$ of FeSe$_{0.67}$Te$_{0.33}$ single crystals we studied in this work. The shadow indicates the phase separation region that has long been considered to exist in bulk FeSe$_{1-x}$Te$_x$.

## 2. Experimental details

FeSe$_{0.67}$Te$_{0.33}$ single crystals were grown by a flux method in two steps [36]. First, polycrystalline FeSe$_{0.67}$Te$_{0.33}$ was prepared by heating stoichiometric mixtures of Fe, Se, and Te powders at 723 K for 24 hours. Next, about 0.3 g of the pre-reacted FeSe$_{0.67}$Te$_{0.33}$ powder and 2.7 g of AlCl$_3$/KCl (molar ration 6:4) were thoroughly mixed by grounding in glove box (O$_2$ and H$_2$O content below 0.1 ppm), and sealed in an evacuated quartz tube. The quartz tube was then placed horizontally in a box furnace, slowly heated to 773 K, maintained for 3 weeks, and then quenched into cold water. Note that, this procedure is different from that in Ref. [36], in which a horizontal two-zone tube furnace was employed. The residual flux was removed by dissolving it in deionized water, and plate-like single crystals with shiny surfaces were finally harvested. An optical image of as-grown single crystals is presented in the inset of figure 2(a). The single crystals were cut into rectangular shapes with dimensions about 600 μm × 400 μm × 17 μm and 380 μm × 320 μm × 30 μm for the transport and magnetization measurements, respectively.

Single-crystal structure was characterized by X-ray diffraction (XRD) measurement using a commercial Rigaku diffractometer with Cu-K$\alpha$ radiation. Elemental analysis was performed by a scanning electron microscope equipped with an energy dispersive X-ray spectroscopy (EDS) probe. Magnetization measurements were performed in a magnetic property measurement system (MPMS-7 T, Quantum Design). Electrical transport measurements were performed in a physical property measurement system (PPMS-9 T, Quantum Design). The Hall resistivity and magnetoresistance (MR) were measured by using a six-lead method with the applied field parallel to the $c$ axis. For the hydrostatic pressure measurements, sample was loaded into a piston-type pressure cell and the actual pressure was determined by measuring the superconducting transition temperatures of lead. Daphne 7373 oil was applied as the pressure transmission medium.

## 3. Results and discussion

3.1 Sample characterizations

Figure 2(a) shows the single-crystal XRD patterns of FeSe$_{0.67}$Te$_{0.33}$ single crystal. Only the (00$l$) diffraction peaks are detected and no trace of a second tetragonal phase formation was found, indicating the good $c$ axis orientation and the absence of tetragonal phase separation in the as-grown single crystals. The $c$ axis lattice constant is calculated as 5.848(3) Å by refining the (00$l$) diffraction peaks, in agreement with the tendency of Te doping in FeSe$_{1-x}$Te$_x$ single crystals [36]. The actual Te concentration was confirmed by the EDS measurement, and a typical EDS spectrum is presented in figure 2(b). Results show that the atomic ration of Fe : Se : Te is 1 : 0.67 : 0.33, which is perfectly consistent with the nominal composition. Compositional mappings in the selected rectangular region of figure 2(c) are shown in figures 2(d)–(f), indicating the homogeneous distribution of Fe, Se, and Te.

The temperature dependence of the in-plane resistivity $\rho_{xx}(T)$ of FeSe$_{0.67}$Te$_{0.33}$ single crystal is shown in figure 2(g). A resistivity anomaly associated with the structural (nematic) transition occurs at $T_s$ = 39 K. $T_s$ is defined as the break point in the plot of d$\rho$/d$T$ versus $T$ shown in the inset, corresponding to the onset point of anomaly in $\rho_{xx}(T)$. The normal state resistivity below $T_s$ can be well described by the equation $\rho_{xx}(T) = \rho_0 + AT^2$, suggesting the Fermi-liquid behavior. This is in stark contrast to that of FeSe$_{1-x}$S$_x$ system, where a $T$-linear dependent resistivity is observed within the nematic phase [16, 17]. With cooling the temperature further, it enters into the superconducting phase, which is more clearly seen in the enlarged view in figure 2(h). The superconducting transition temperature is marked by $T_c^{onset}$, $T_c^{middle}$, and $T_c^{zero}$, corresponding to 9.8 K, 8.2 K, and 6.2 K, respectively. Moreover, the superconducting transition can also be seen in the temperature dependence of zero-field-cooled (ZFC) magnetization, as shown in the inset. Clearly, a superconducting diamagnetic signal at $T_c^M = 8.2$ K is detected, consistent with the resistivity measurement.

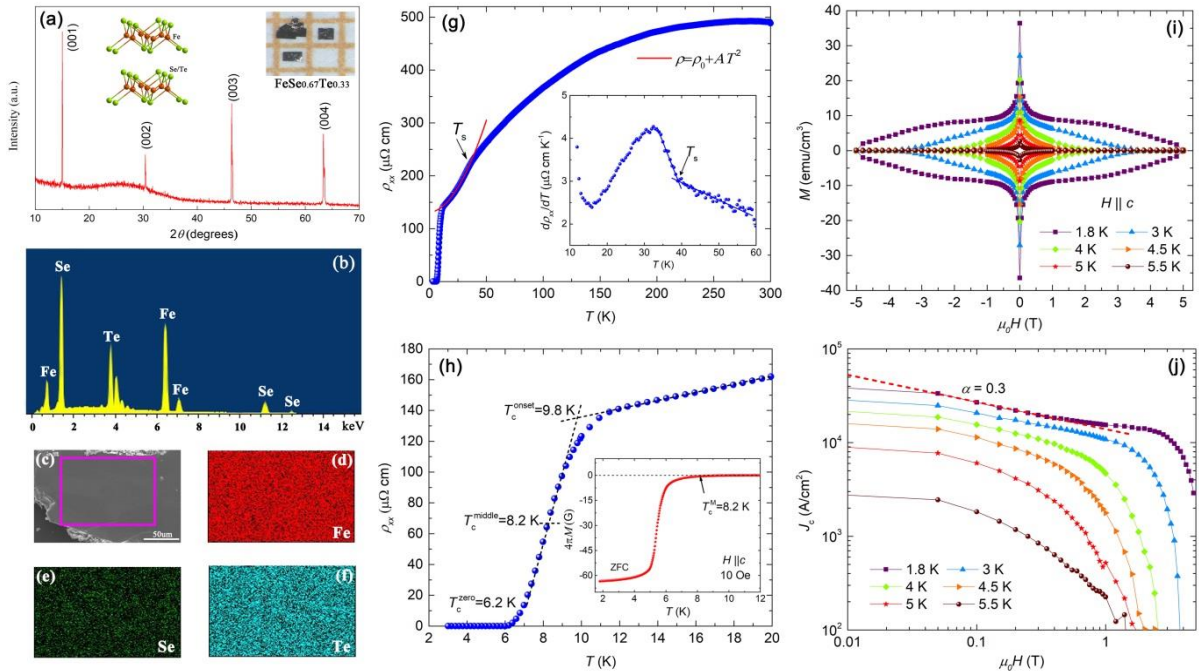

**Figure 2.** (a) Single-crystal XRD patterns of FeSe$_{0.67}$Te$_{0.33}$ single crystal. The inset shows the crystal structure (left) and optical image of the obtained single crystals (right). Chemical composition analysis: (b) A typical EDS spectrum taken on the selected rectangular region in SEM image (c), and corresponding EDS mapping images for each element (d) Fe, (e) Se, and (f) Te. (g) Temperature dependence of the in-plane resistivity $\rho_{xx}(T)$ of FeSe$_{0.67}$Te$_{0.33}$ single crystal. The inset shows the temperature derivative of $\rho_{xx}(T)$, the nematic transition temperature $T_s$ is defined as the break point as indicated by the arrow. The red curve below $T_s$ is the fit to Fermi-liquid formula $\rho_{xx}(T) = \rho_0 + AT^2$. (h) Enlarged view of $\rho_{xx}(T)$ near the superconducting transition. Inset: temperature dependence of the ZFC magnetization with an applied field of 10 Oe parallel to the $c$ axis. (i) The MHLs of FeSe$_{0.67}$Te$_{0.33}$ single crystal at various temperatures ranging from 1.8 to 5.5 K. (j) Temperature dependence of $J_c$ derived from Bean model. The red dashed line represents the power-law decay of $H^{-\alpha}$.

Figure 2(i) presents the magnetization hysteresis loops (MHLs) of the FeSe$_{0.67}$Te$_{0.33}$ single crystal at various temperatures for $H \parallel c$ axis. The symmetric MHLs indicate the dominance of the bulk pinning. We calculated the field dependence of $J_c$ from the MHLs based on the Bean model $J_c = 20\Delta M/[a(1 - a/3b)]$, where $\Delta M$ is $M_{down}-M_{up}$, $M_{up}$ [emu/cm$^3$] and $M_{down}$ [emu/cm$^3$] are the magnetization when sweeping fields up and down, respectively, $a$ [cm] and $b$ [cm] are sample widths

($a < b$) [39]. The derived $J_c$ as a function of magnetic field is shown in figure 2(j). $J_c$ reaches a value of 6.3×10$^4$ A/cm$^2$ at zero filed ($T$ = 1.8 K), which is slightly higher than that of FeSe single crystal [40], but is still smaller than that of highly Te-doped ones [41, 42]. The value of $J_c$ changes only slightly with increasing field below 500 Oe, followed by a power-law dependence ($J_c \propto H^{-\alpha}$) related to strong pinning centers. Such power law dependence of $J_c$ with $\alpha$ = 0.3 was also observed in our H$^+$-irradiated FeSe [43], Ba(Fe$_{0.93}$Co$_{0.07}$)$_2$As$_2$ [44], and Ba$_{0.6}$K$_{0.4}$Fe$_2$As$_2$ single crystals [45].

3.2 Upper critical field

Figures 3(a) and (b) present the temperature dependence of normalized resistivity under different magnetic fields ranging from 0 to 9 T for $H \parallel c$ axis and $H \parallel ab$ plane, respectively. As the applied field increases, the resistivity transition width becomes slightly broader and the onset of SC gradually shifts to lower temperatures. This trend is more pronounced for $H \parallel c$ axis than $H \parallel ab$ plane, in agreement with that in FeSe$_{1-x}$Te$_x$ single crystals with higher Te content [46]. To minimize the influence of superconducting fluctuations and vortex motion in the vortex-liquid phase, we adopt the midpoint of superconducting transition to determine the $\mu_0 H_{c2}$, which is presented in figure 3(c) for both directions. The slopes of $\mu_0 H_{c2}$ near $T_c$, $\mu_0 dH_{c2}/dT$, are -3.6 T/K and -7.3 T/K for $H \parallel c$ axis and $H \parallel ab$ plane, respectively. According to the Werthamer-Helfand-Hohenberg (WHH) theory [47], the orbital limited upper critical field, $\mu_0 H_{c2,orb}(0)$, can be described by

$$\mu_0 H_{c2,orb}(0) = -0.693 T_c \left(\frac{\mu_0 dH_{c2}}{dT}\right)_{T=T_c}. \quad (1)$$

Taking $T_c^{\text{middle}}$ =8.2 K, the values of $\mu_0 H_{c2,orb}^{ab}(0)$ and $\mu_0 H_{c2,orb}^{c}(0)$ are estimated as 41.5 T and 20.5 T, respectively. Using the Ginzburg-Landau (GL) formula, $\mu_0 H_{c2,orb}^{c}(0)=\Phi_0/2\pi\xi_{ab}^2(0)$ and $\mu_0 H_{c2,orb}^{ab}(0)=\Phi_0/2\pi\xi_{ab}(0)\xi_c(0)$, where $\Phi_0 = 2.07\times10^{-15}$ T m$^2$ is the flux quantum, the GL coherence lengths are obtained as $\xi_{ab}(0) = 4.0$ nm and $\xi_c(0) = 2.0$ nm.

Generally, there are two distinct mechanisms that act on the suppression of SC under magnetic fields in type-II superconductors [46, 48, 49]. One is the orbital pair-breaking effect, with opposite momenta acting on the paired electrons. In this case, the SC is destroyed when the kinetic energy of the Cooper pairs exceeds the condensation energy. The other is attributed to the spin-paramagnetic pair-breaking effect, which comes from the Zeeman splitting of spin singlet cooper pairs. The SC is also eliminated when the Pauli spin susceptibility energy is larger than the condensation energy.

To quantitatively describe our results, the full WHH formula including the effects of spin-paramagnetism (via the Maki parameter $\alpha$) and spin-orbit interaction (via $\lambda_{so}$), is used to fit the $\mu_0 H_{c2}(T)$ data. According to the WHH theory [47], $\mu_0 H_{c2}(T)$ in the dirty limit can be described by the digamma function

$$\ln\frac{1}{t} = \left(\frac{1}{2} + \frac{i\lambda_{so}}{4\gamma}\right)\Psi\left(\frac{1}{2} + \frac{\bar{h}+\lambda_{so}/2+i\lambda}{2t}\right) + \left(\frac{1}{2} - \frac{i\lambda_{so}}{4\gamma}\right)\Psi\left(\frac{1}{2} + \frac{\bar{h}+\lambda_{so}/2-i\lambda}{2t}\right) - \Psi\left(\frac{1}{2}\right), \quad (2)$$

where $t = T/T_c$, $\gamma \equiv \left[(\alpha\bar{h})^2 - (\lambda_{so}/2)^2\right]^{1/2}$ and $\bar{h} = \frac{4\bar{h}}{\pi^2(-d\bar{h}/dt)_{t=1}} = \frac{4H_{c2}}{\pi^2(-dH_{c2}/dt)_{t=1}}$. In the condition of the absence of both spin-paramagnetic effect and spin-orbit interaction, $\alpha$=0 and $\lambda_{so}$=0, the orbital-limited upper critical field can be expressed as Eq. (1) as mentioned above.

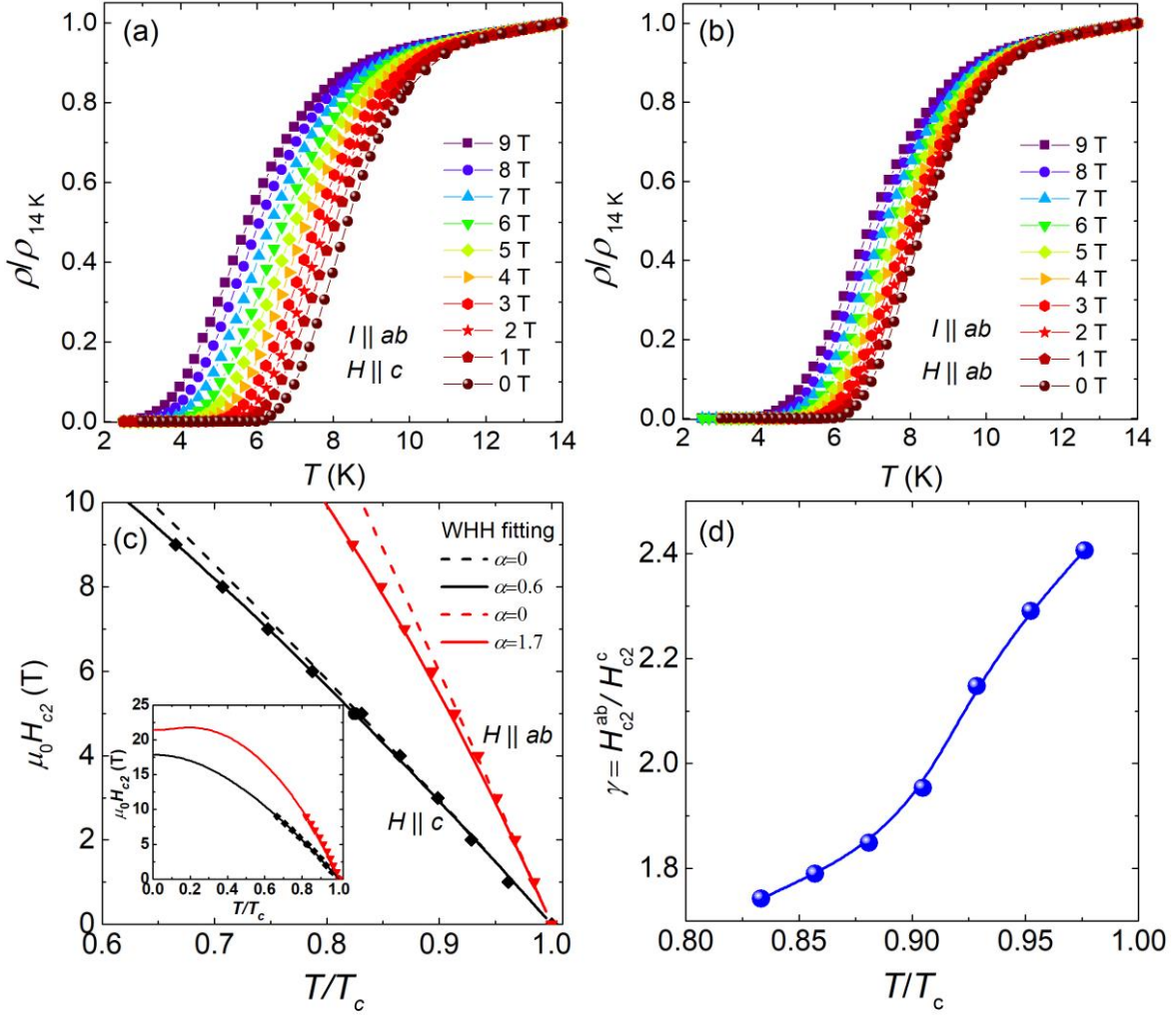

**Figure 3.** Temperature dependence of normalized resistivity under various magnetic fields up to 9 T for (a) $H \parallel c$ axis and (b) $H \parallel ab$ plane. (c) Temperature dependence of $\mu_0 H_{c2}$ for both field directions extracted from panels (a) and (b) using the criterion of 50% $\rho_n$. The dashed (solid) curves represent the WHH model fitting without (with) considering the spin-paramagnetic effect (see text for detail). Inset shows the WHH fitting curves over the full temperature range below $T_c$. (d) The anisotropy of $\mu_0 H_{c2}(T)$, $\gamma = H_{c2}^{ab}(T)/H_{c2}^{c}(T)$, as a function of reduced temperature.

As clearly seen from figure 3(c), the data of $\mu_0 H_{c2}(T)$ for both field directions fall below the prediction of WHH model without considering the spin paramagnetic effect, i.e., $\alpha = 0$ (dotted lines), indicating that the spin-paramagnetic effect should be considered to reproduce the experimental data. By properly adjusting the value of $\alpha$, the best fits were obtained with $\alpha = 1.7$ and 0.6 for $H \parallel ab$ plane and $H \parallel c$ axis, respectively, as shown by the solid lines. It is noteworthy here that the spin-orbit scattering is not necessary to have the best fit ($\lambda_{so} = 0$). According to the Maki formula [50], $\alpha$ can be expressed as $\alpha = \sqrt{2} H_{c2}^{orb}(0)/H_p(0)$, where $\mu_0 H_p(0)$ is the zero temperature Pauli limited field. In the weak-coupling BCS superconductors, the Pauli limited field is given by $\mu_0 H_p(0) = \mu_0 H_p^{BCS}(0) = 1.84\, T_c$, yielding the value of $\mu_0 H_p^{BCS}(0) = 15$ T for FeSe$_{0.67}$Te$_{0.33}$ single crystal. Thus, the values of $\alpha$ calculated from the Maki formula are 3.9 for $H \parallel ab$ plane and 1.9 for $H \parallel c$ axis, which are 2.3~3 times larger than those of fitting results. Such deviation has been widely observed in IBSs, which is attributed to the enhancement of $\mu_0H_p(0)$ over $\mu_0H_p^{BCS}(0)$ due to the strong coupling or correlation effects [49].

In FeSe single crystals, for $H \parallel ab$ plane, $\mu_0H_{c2}(T)$ shows an anomalous upturn at low temperatures below $T^*$, which is suggested to be associated with inherent SDW instability of quasiparticles or a possible Fulde-Ferrel-Larkin-Ovchinnikov (FFLO) state [51, 52]. Above $T^*$, the behavior of $\mu_0H_{c2}(T)$ can be well described by the WHH model with $\alpha = 1.5$ [52], which is comparable to that of FeSe$_{0.67}$Te$_{0.33}$ in this work. While, for $H \parallel c$ axis, $\mu_0H_{c2}(T)$ shows sublinear temperature dependence with decreasing temperature and a two-band model is necessary to reproduce the data [52]. Therefore, our result suggests that the two-band effect on $\mu_0H_{c2}(T)$ is gradually suppressed by Te doping and spin-paramagnetic effect becomes dominant instead for $H \parallel c$ axis. Indeed, the dominance of spin-paramagnetic effect in both field directions was also observed in previous higher Te doped single crystals, e.g., FeSe$_{0.39}$Te$_{0.61}$ and FeSe$_{0.11}$Te$_{0.89}$ [46].

Figure 3(d) shows the anisotropy of $\mu_0H_{c2}(T)$, $\gamma = H_{c2}^{ab}(T)/H_{c2}^{c}(T)$, as a function of reduced temperature $t=T/T_c$. The magnitude of $\gamma$ decreases gradually with decreasing temperature, residing in the range of 1.7-2.4. Previous study has revealed that the $\gamma$ value of FeSe$_{1-x}$Te$_x$ single crystal with high-Te concentration decreases toward 1 at low temperatures, indicative of a high-field isotropic superconductor [32, 46]. Certainly, high field study is needed to further clarify the behavior of $\mu_0H_{c2}(T)$ in FeSe$_{1-x}$Te$_x$ single crystals in the region of $0 < x \leq 0.5$.

3.3 Hall effect and magnetoresistance

It is well known that the transport properties in the normal state, such as Hall effect and MR, can reflect the electronic scattering and the topology of Fermi surface, which is the key step to understand the superconducting mechanism. The inset of figure 4(a) shows the magnetic field dependence of Hall resistivity $\rho_{yx}$ at different temperatures for FeSe$_{0.67}$Te$_{0.33}$ single crystal. In all cases, $\rho_{yx}$ exhibits good linear field dependence up to 9 T. The Hall coefficient $R_H$ determined by $R_H = \rho_{yx}/H$ is plotted as a function of temperature in the main panel of figure 4(a). $R_H$ keeps nearly temperature independent at high temperatures, followed by a gradual decrease with decreasing temperature below 175 K. It is noted that a sign reversal from positive to negative occurs at around 80 K, indicating that the dominant carriers change from hole-type to electron-type. Furthermore, $R_H$ decreases more rapidly when temperature is decreased below $T_s$, signifying the Fermi surface reconstruction induced by the structural transition as observed in FeSe [53]. In a single-band conventional metal with Fermi liquid behavior, $R_H$ is independent of temperature. However, $R_H$ varies with temperature for a multiband material, e.g., MgB$_2$ [54] and many IBSs [55-58], or a sample with non-Fermi liquid behavior resulted from the anisotropic scattering due to spin fluctuations, e.g., high-$T_c$ cuprates [59]. Therefore, the strong temperature dependence of $R_H$ and Fermi-liquid behavior at low temperatures indicate the multiband feature in FeSe$_{0.67}$Te$_{0.33}$ single crystal.

To gain more insight into the magneto-transport properties, we also performed the MR measurement on the same crystal with $H \parallel c$ axis. Figure 4(b) shows the magnetic field dependence of MR, $\Delta\rho_{xx}(H)/\rho_{xx}(0) = [\rho_{xx}(H)-\rho_{xx}(0)]/\rho_{xx}(0)$, at temperatures ranging from 15 to 40 K. The magnitude of

MR is only about 1% at 15 K and 9 T, which is much smaller than that in FeSe [56] and fully annealed FeSe$_{0.4}$Te$_{0.6}$ [55] single crystals. As temperature increases, MR gradually decreases and becomes almost negligible for $T$ above 40 K.

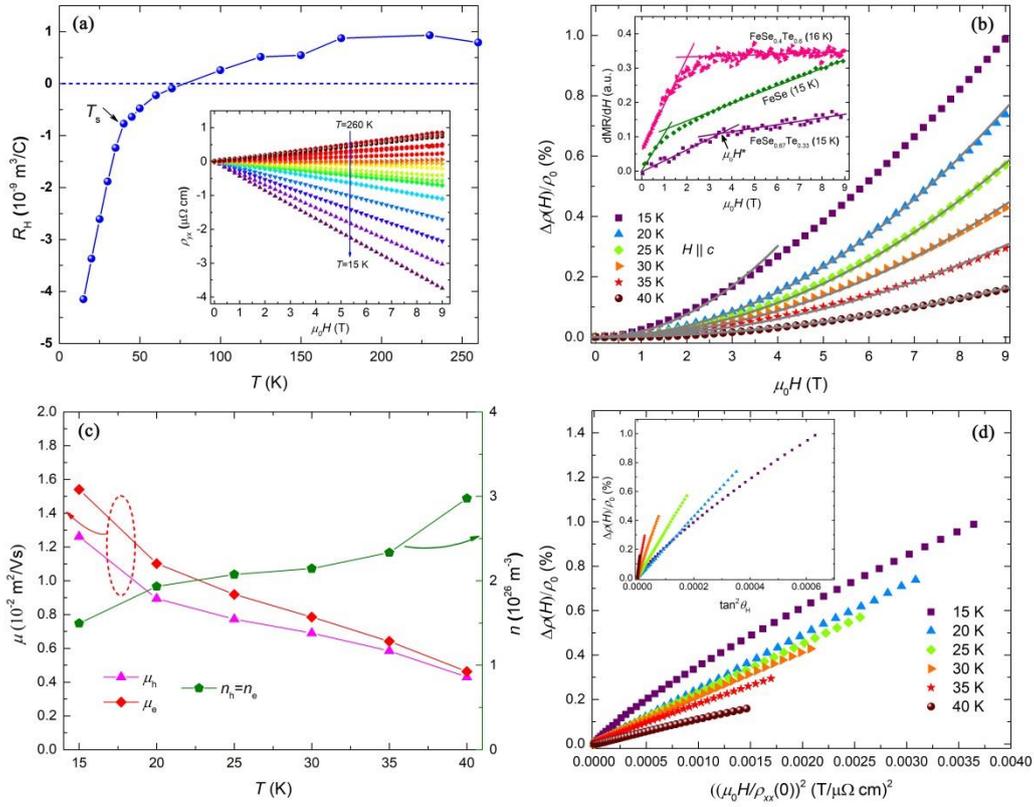

**Figure 4.** (a) Hall coefficient $R_H$ of FeSe$_{0.67}$Te$_{0.33}$ single crystal as a function of temperature. The inset shows the magnetic field dependence of Hall resistivity $\rho_{yx}$ at different temperatures ranging from 260 to 15 K. (b) Magnetic field dependence of MR at low temperatures ($T$ = 15, 20, 25, 30, 35, and 40 K) with $H \parallel c$ axis. The gray curves represent the $(\mu_0H)^2$ fitting. Inset: the field derivative of MR, d(MR)/d($\mu_0H$) at $T$ = 15 K. Corresponding data of FeSe single crystal at $T$ = 15 K and FeSe$_{0.4}$Te$_{0.6}$ single crystal at $T$ = 16 K reported in Refs. [55, 56] are also included for comparison. The arrow indicates the critical field $\mu_0H^*$, defined as the intercept point of two fitting lines in low field semi-classical regime and high field quantum regime. (c) Temperature dependence of carrier density ($n_e = n_h$) and mobility of electrons ($\mu_e$) and holes ($\mu_h$) derived using a compensated two-band model (see text for details). (d) MR plotted as a function of $(\mu_0H/\rho_{xx}(0))^2$ (main panel) and $\tan^2\theta_H$ (inset).

Intriguingly, the MR at $T$ = 15 K reveals a crossover from a semi-classical quadratic dependence at low fields to a more linear behavior at high fields across a critical field $\mu_0H^*$. This behavior can be seen more clearly from the field derivative of d(MR)/d($\mu_0H$), as shown in the inset of figure 4(b). d(MR)/d($\mu_0H$) linearly increases with magnetic field in the low field region and MR exhibits a semi-classical $(\mu_0H)^2$ dependence. However, above $\mu_0H^*$, the slope of d(MR)/d($\mu_0H$) suddenly reduces, indicating that the MR follows linear field dependence plus a quadratic term [56]. The linear MR has been observed in our previous studies on FeSe and FeSe$_{0.4}$Te$_{0.6}$ single crystals [55, 56], which was ascribed to the existence of Dirac-cone state, as confirmed by angle-resolved photoemission spectroscopy (ARPES) measurement [27, 29, 60]. For reference, typical data of d(MR)/d($\mu_0H$) for FeSe and FeSe$_{0.4}$Te$_{0.6}$ single crystals at a given temperature $T$ = 15 K or 16 K are also included in the inset of figure 4(b). As we discussed previously, linear MR may come from the quantum effects when all carriers degenerate into the lowest Landau level (LL), i.e., in the quantum limit regime [55, 56, 61]. Quantum limit can be readily achieved in a Dirac system by a moderate field as its zeroth and first LLs are well separated. Thus, the occurrence of $\mu_0H^*$ in FeSe$_{0.67}$Te$_{0.33}$ single crystal indicates that the Dirac-cone state might also exist. However, for $T > 15$ K, the $\mu_0H^*$ cannot be easily distinguished, due to the smaller MR signal at high temperatures.

In a system with multiband electronic structure, the most natural approach to analyse these transport properties is the two-band model (i.e., one with electron and the other with hole), and the resistivity tensor can be expressed as

$$\rho_{xx}(0) = \frac{1}{e}\frac{1}{(\mu_e n_e + \mu_h n_h)} \quad\quad\quad\quad\quad\quad\quad\quad\quad\quad\quad (3)$$

$$\rho_{yx}(H) = \frac{1}{e}\frac{(\mu_h^2 n_h - \mu_e^2 n_e) + (\mu_h \mu_e)^2 (\mu_0 H)^2 (n_h - n_e)}{(\mu_e n_e + \mu_h n_h)^2 + (\mu_h \mu_e)^2 (\mu_0 H)^2 (n_h - n_e)^2}\mu_0 H \quad\quad (4)$$

$$\frac{\Delta\rho_{xx}(H)}{\rho_{xx}(0)} = \frac{n_h \mu_h n_e \mu_e (\mu_h + \mu_e)^2 (\mu_0 H)^2}{(\mu_e n_e + \mu_h n_h)^2 + (\mu_h \mu_e)^2 (\mu_0 H)^2 (n_h - n_e)^2}, \quad\quad\quad (5)$$

where $n_e$ ($n_h$) and $\mu_e$ ($\mu_h$) denote the electron (hole) density and electron (hole) mobility, respectively [62]. Note that $\rho_{yx}(H)$ is linear in field up to 9 T, i.e., the nonlinearity term $(n_h-n_e)(\mu_0H)^2$ is negligible, indicating that FeSe$_{0.67}$Te$_{0.33}$ is a compensated semimetal with almost equal number of electrons and holes, $n = n_e \approx n_h$. Therefore, Eqs. (3)-(5) can be simplified as $\rho_{xx}(0) = \frac{1}{ne}\frac{1}{\mu_e+\mu_h}$, $\rho_{yx}(H) = \frac{1}{ne}\frac{\mu_h-\mu_e}{\mu_h+\mu_e}\mu_0 H$, and $\frac{\Delta\rho_{xx}(H)}{\rho_{xx}(0)} = \mu_h\mu_e(\mu_0H)^2$, respectively. The quadratic dependence of MR is consistent with our results as shown in figure 4(b). This compensated two-band model has been used for LiFeAs [57], BaFe$_2$(As$_{1-x}$P$_x$)$_2$ [63], and FeSe [64] single crystals. Herein, we adopt it to fit the data of the MR and $\rho_{yx}(H)$ in FeSe$_{0.67}$Te$_{0.33}$ single crystal. At 15 K, the fitting is performed on the low field part below $\mu_0H^*$ to avoid the influence of linear MR at high field. The obtained $T$ variations of $n$, $\mu_e$, and $\mu_h$ are presented in figure 4 (c). The carrier density $n$ is about $3\times10^{26}$ m$^{-3}$ at 40 K, which is comparable to that of FeSe$_{0.7}$Te$_{0.3}$ thin films [65]. With decreasing temperature, $n$ is gradually reduced, whereas, both $\mu_e$ and $\mu_h$ increase. At $T_s$, the difference between $\mu_e$ and $\mu_h$ is small, but it increases with decreasing temperature, resulting in a more negative $R_H$ below $T_s$.

According to the semiclassical Boltzmann transport theory, in the Fermi liquid state of a single-band system with isotropic scattering, the MR at different temperatures can be scaled by the Kohler's rule [66, 67], $\Delta\rho_{xx}(H)/\rho_{xx}(0)=F(H\tau)$, where $\tau$ is the scattering time. Since $\rho_{xx}(0)=m^*/ne^2\tau$, where $m^*$ is the effective mass and $n$ is the carrier density, the Kohler's rule can be rewritten as $\Delta\rho_{xx}(H)/\rho_{xx}(0)=F[Hm^*/\rho_{xx}(0)ne^2]$. If the factor $m^*/ne^2$ is not sensitive to temperature, the Kohler's rule can be simplified as the common form $\Delta\rho_{xx}(H)/\rho_{xx}(0)=f[H/\rho_{xx}(0)]$ [66]. The departure from the Kohler's scaling is generally believed to result from the change of carrier density with temperature or the anisotropic scattering rate. To examine Kohler's rule, the MR as a function of $[\mu_0H/\rho_{xx}(0)]^2$ is shown in figure 4(d). Clearly, the MR data measured at different temperatures do not collapse into a single curve, i.e., the Kohler's rule is not obeyed. In cuprates and IBSs, the violation of the Kohler's rule has been observed in the vicinity of AFM phase, which is attributed to the existence of hot and cold spots on the Fermi surface with anisotropic scattering rate due to the AFM fluctuations [63, 68, 69]. In these compounds, the resistivity generally exhibits a non-Fermi-liquid behavior and MR can be scaled by the modified Kohler's rule, $\Delta\rho_{xx}(H)/\rho_{xx}(0) \propto \tan^2\theta_H = [\rho_{yx}(H)/\rho_{xx}(0)]^2$ [63, 68, 69]. We also plot $\Delta\rho_{xx}(H)/\rho_{xx}(0)$ as a function of $\tan^2\theta_H$ in the inset of figure 4(d). Obviously, the modified Kohler's rule is also violated. Thus, the invalidity of modified Kohler's rule as well as the Fermi-liquid ground state in $FeSe_{0.67}Te_{0.33}$ single crystal suggest that the anisotropic scattering caused by AFM fluctuations is unlikely to be responsible for the violation of the Kohler's rule. Actually, in the multiband system, $\tau$ is strongly band dependent, and its temperature dependent behavior also varies from one band to another. So, the single-band Kohler's rule is found to be violated in many multiband materials such as $MgB_2$ [70] and some IBSs [55, 58, 71]. In $FeSe_{0.67}Te_{0.33}$ single crystal, given the strong temperature dependence of $R_H$ and the different temperature dependent behavior of mobility for electrons and holes revealed by the two-band analysis (see figures 4 (a) and (c)), we believe that the violation of the (modified) Kohler's rule should also be resulted from the multiband effects, similar to $FeSe_{0.4}Te_{0.6}$ and FeS single crystals [55, 58].

3.4 Hydrostatic pressure effect

Recently, high pressure measurements have been widely performed on FeSe and $FeSe_{1-x}S_x$ single crystals, which reveal a complex and interesting phase interplay [10-14, 72-76]. Herein, we also performed hydrostatic pressure studies on $FeSe_{0.67}Te_{0.33}$ single crystals. The $\rho_{xx}(T)$ curves of $FeSe_{0.67}Te_{0.33}$ single crystal at different pressures ranging from 0 to 2.31 GPa are shown in figure 5(a). Upon applying pressure, the nematic transition is quickly suppressed and becomes invisible above $P = 0.29$ GPa. Besides, $T_c$ is monotonically increased with pressure, reaching $T_c^{zero} = 25$ K at $P = 2.31$ GPa, the highest pressure of the present experiment. The simultaneous suppression of $T_s$ and enhancement of $T_c$ indicate the competition between nematicity and superconductivity. The superconducting transition becomes much sharper under high pressure, suggesting that the superconducting phase becomes quite homogeneous at high pressures. Based on these resistivity measurements, the T-P phase diagram of $FeSe_{0.67}Te_{0.33}$ single crystals is constructed in figure 5(b), in which $T_s$ and $T_c^{zero}$ were obtained using the criteria as mentioned above (see figure 2).

It is noted that in the preliminary pressure study of polycrystalline $FeSe_{0.57}Te_{0.43}$ [77], high-resolution synchrotron XRD experiments revealed a structural transition from tetragonal to orthorhombic phase at ambient pressure at $T_s \sim 40$ K, almost consistent with $FeSe_{0.67}Te_{0.33}$ single crystal. By the application of pressure, the orthorhombic phase in polycrystalline $FeSe_{0.57}Te_{0.43}$ can survive up to 2.5 GPa. This is distinctly different from $FeSe_{0.67}Te_{0.33}$ single crystal, where $T_s$ is quickly suppressed with pressure. We suspect that this discrepancy could be due to the partially inhomogeneous distribution of the Te dopant in polycrystalline sample. Indeed, the $T_s$ value in polycrystalline

FeSe$_{0.57}$Te$_{0.43}$ is much higher than that reported on single crystal (see the phase diagram in figure 1). Also, no anomaly associated with structural transition was observed in the resistivity of polycrystalline sample [77].

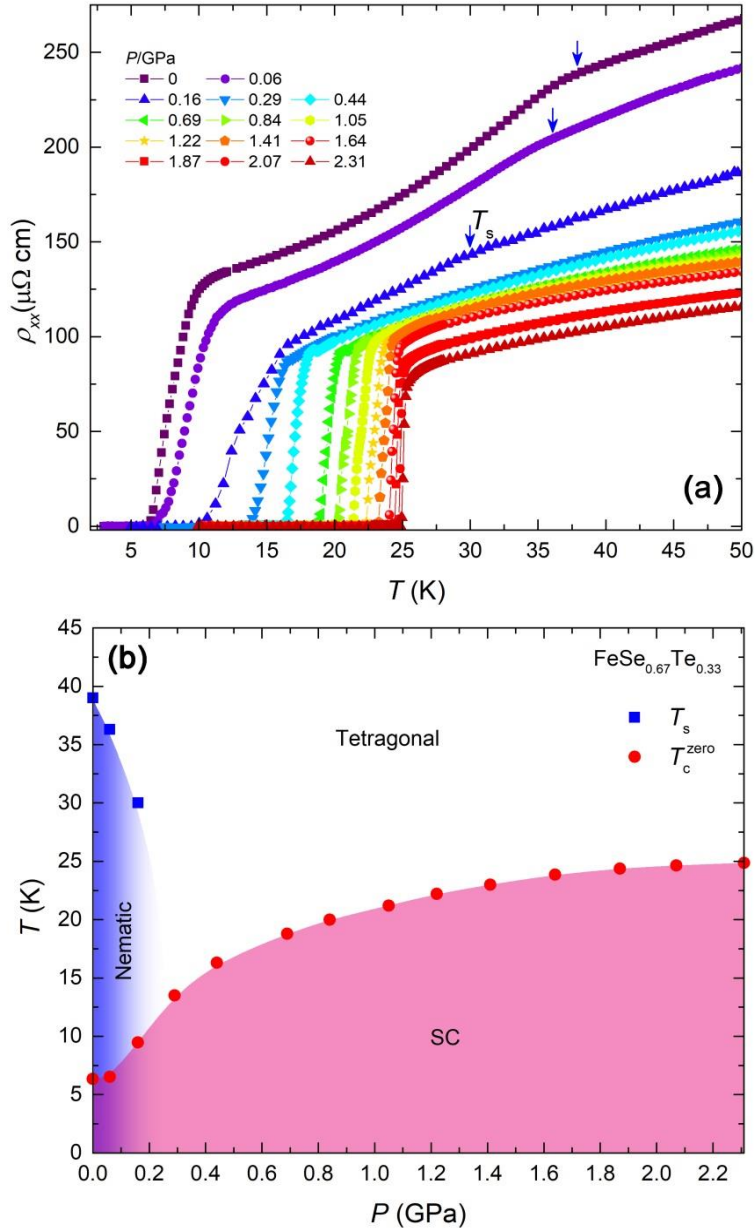

**Figure 5.** (a) Evolution of $\rho_{xx}(T)$ with hydrostatic pressure up to 2.31 GPa for FeSe$_{0.67}$Te$_{0.33}$ single crystal. (b) $T$-$P$ phase diagram determined from the resistivity measurements in panel (a).

Next, we compare the phase diagram to that of FeSe$_{1-x}$S$_x$ single crystals. It has been found that the magnetic order in FeSe$_{1-x}$S$_x$ ($0 \leq x \leq 0.14$) single crystals can be stabilized inside the nematic phase in the low-pressure region, which competes with SC, leading to a local maximum (small dome) in $T_c$, where the magnetic order arises [72, 73]. The local maximum of $T_c$ shifts to lower pressure with increasing S content. However, in FeSe$_{0.67}$Te$_{0.33}$ single crystals, $T_c$ increases monotonically with pressure up to 2.31 GPa, and no local maximum of $T_c$ accompanied by the formation of magnetic order is observed. On the other hand, another prominent dome of magnetic order was also observed in the higher-pressure range, detached from the nematic phase for $x \geq 0.04$ [75]. It indicates that two magnetic phases exist in FeSe$_{1-x}$S$_x$ single crystals, which increasingly separate for higher $x$ [72, 73]. Matsuura *et al*. have proposed that the increase in chalcogen height $h_{Ch}$ may be responsible for the occurrence of magnetic order in FeSe$_{1-x}$S$_x$ single crystals in the high-pressure region [75]. In contrast to the physical pressure results (i.e., increasing $h_{Ch}$), $h_{Ch}$ decreases monotonically upon S doping. As a consequence, the dome-shaped magnetic order shifts to higher pressures with increasing S concentration, as higher pressures are required for obtaining a larger $h_{Ch}$ to induce magnetic order in FeSe$_{1-x}$S$_x$ single crystals [75]. Considering that the ionic radius of Te is larger than that of Se, the effects on the structural parameters should be opposite to that of S substitution, i.e., increasing $h_{Ch}$ with Te substitution [78]. In this regard, the pressure effects in the FeSe$_{1-x}$Te$_x$ are expected to be different from those in the FeSe$_{1-x}$S$_x$. Further efforts are expected to clarify how the magnetic order evolves with pressure and Te substitution in the higher-pressure region in FeSe$_{1-x}$Te$_x$ single crystals, which will gain more insights into the origin of magnetic order, as well as the complicated interplay among different quantum orders in iron chalcogenide superconductors.

## 4. Summary

In summary, we have grown FeSe$_{0.67}$Te$_{0.33}$ single crystals without phase separation by a flux method, and comprehensively studied the electronic transport properties and hydrostatic pressure effect. FeSe$_{0.67}$Te$_{0.33}$ undergoes a nematic transition at $T_s = 39$ K, below which resistivity exhibits a Fermi-liquid behaviour, in contrast to the FeSe$_{1-x}$S$_x$ system with linear temperature dependence. Analysis based on the WHH model indicates that $\mu_0H_{c2}(T)$ in both field directions are affected by the spin-paramagnetic effect. The MR reveals a crossover from the low-$H$ quadratic to the high-$H$ quasi-linear behavior at a critical field, signifying the possible existence of Dirac-cone state. Besides, the strong temperature dependence of Hall coefficient, violation of (modified) Kohler's rule, and two-band model analysis suggest the multiband effects in FeSe$_{0.67}$Te$_{0.33}$ single crystals. Hydrostatic pressure study reveals that $T_s$ is quickly suppressed while $T_c$ is monotonically increased under pressure, indicating the competition between nematicity and superconductivity.

**Acknowledgements**

This work was partly supported by the National Key R&D Program of China (Grant No. 2018YFA0704300), the Strategic Priority Research Program (B) of the Chinese Academy of Sciences (Grant No. XDB25000000), the National Natural Science Foundation of China (Grants No. U1932217 and 11674054), and the Fundamental Research Funds for the Central Universities. X. Z. Xing was also supported by a project funded by the China Postdoctoral Science Foundation (Grant No. 2019M661679) and the Jiangsu Planned Projects for Postdoctoral Research Funds (Grant No. 2019K149). Y. Sun was supported by JSPS KAKENHI (Grant Nos. JP20H05164 and JP19K14661). T. Tamegai was supported by JSPS KAKENHI (Grant No. 17H01141).

*Note added in proof.* After the completion of this work, we became aware of a recent high prssure study of FeSe$_{1-x}$Te$_x$ ($0.04 \leq x \leq 0.5$) single crystals [79], which reveals that, in contrast to FeSe$_{1-x}$S$_x$

[75], enhanced superconductivity in FeSe$_{1-x}$Te$_x$ does not correlate with magnetism but with the suppression of nematicity, signifying the important role of nematic fluctuations for high-temperature SC in FeSe$_{1-x}$Te$_x$ system. The *T-P* phase diagrams of FeSe$_{1-x}$Te$_x$ (0.14 ≤ *x* ≤ 0.38) in the low pressure region are consistent with the present study for *x*=0.33.

**Reference**

[1] Hsu F C *et al* 2008 *Proc. Natl. Acad. Sci. USA* **105** 14262.

[2] Shibauchi T, Hanaguri T, and Matsuda Y 2020 *J. Phys. Soc. Jpn.* **89** 102002.

[3] S. Medvedev *et al* 2009 *Nat. Mater.* **8** 630.

[4] Ge J F, Liu Z L, Liu C, Gao C L, Qian D, Xue Q K, Liu Y, and Jia J F 2015 *Nat. Mater.* **14** 285-289.

[5] Wang Q-Y *et al* 2012 *Chin. Phys. Lett.* **29** 037402.

[6] Kasahara S *et al* 2014 *Proc. Natl. Acad. Sci. USA* **111** 16309-16313.

[7] McQueen T M, Williams A J, Stephens P W, Tao J, Zhu Y, Ksenofontov V, Casper F, Felser C, and Cava R J 2009 *Phys. Rev. Lett.* **103** 057002.

[8] Baek S H, Efremov D V, Ok J M, Kim J S, van den Brink J, and Buchner B 2015 *Nat. Mater.* **14** 210.

[9] Bohmer A E, Arai T, Hardy F, Hattori T, Iye T, Wolf T, Lohneysen H V, Ishida K, and Meingast C 2015 *Phys. Rev. Lett.* **114** 027001.

[10] Bendele M, Amato A, Conder K, Elender M, Keller H, Klauss H H, Luetkens H, Pomjakushina E, Raselli A, and Khasanov R 2010 *Phys. Rev. Lett.* **104** 087003.

[11] Bendele M, Ichsanow A, Pashkevich Y, Keller L, Strässle T, Gusev A, Pomjakushina E, Conder K, Khasanov R, and Keller H 2012 *Phys. Rev. B* **85** 064517.

[12] Sun J P *et al* 2016 *Nat. Commun.* **7** 12146.

[13] Wang P S, Sun S S, Cui Y, Song W H, Li T R, Yu R, Lei H, and Yu W 2016 *Phys. Rev. Lett.* **117** 237001.

[14] Terashima T *et al* 2015 *J. Phys. Soc. Jpn.* **84** 063701.

[15] Coldea A I arXiv:2009.05523.

[16] Bristow M, Reiss P, Haghighirad A A, Zajicek Z, Singh S J, Wolf T, Graf D, Knafo W, McCollam A, and Coldea A I 2020 *Phys. Rev. Research* **2** 013309.

[17] Licciardello S, Buhot J, Lu J, Ayres J, Kasahara S, Matsuda Y, Shibauchi T, and Hussey N E 2019 *Nature* **567** 213-217.

[18] Sato Y, Kasahara S, Taniguchi T, Xing X, Kasahara Y, Tokiwa Y, Yamakawa Y, Kontani H, Shibauchi T, and Matsuda Y 2018 *Proc. Natl. Acad. Sci. USA* **115** 1227-1231.

[19] T. Hanaguri K I, Y. Kohsaka, T. Machida, T. Watashige, S. Kasahara, T. Shibauchi, and Y. Matsuda 2018 *Sci. Adv.* **4** eaar6419.

[20] Coldea A I *et al* 2019 *npj Quan. Mater.* **4** 2.

[21] Yi X, Xing X, Qin L, Feng J, Li M, Zhang Y, Meng Y, Zhou N, Sun Y, and Shi Z arXiv:2010.05191.

[22] Mizuguchi Y, and Takano Y 2010 *J. Phys. Soc. Jpn.* **79** 102001.

[23] Yeh K-W *et al* 2008 *EPL (Europhysics Letters)* **84** 37002.

[24] Sun Y, Tsuchiya Y, Taen T, Yamada T, Pyon S, Sugimoto A, Ekino T, Shi Z, and Tamegai T 2014 *Sci. Rep.* **4** 4585.

[25] Sun Y, Kittaka S, Nakamura S, Sakakibara T, Irie K, Nomoto T, Machida K, Chen J, and Tamegai T 2017 *Phys. Rev. B* **96** 220505(R).

[26] Hanaguri T, Niitaka S, Kuroki K, and Takagi H 2010 *Science* **328** 474.

[27] Zhang P *et al* 2018 *Science* **360** 182.

[28] Miao H *et al* 2012 *Phys. Rev. B* **85** 094506.

[29] Zhang P *et al* 2018 *Nat. Phys.* **15** 41-47.

[30] Machida T, Sun Y, Pyon S, Takeda S, Kohsaka Y, Hanaguri T, Sasagawa T, and Tamegai T 2019 *Nat. Mater.* **18** 811-815.

[31] Wang D *et al* 2018 *Science* **362** 333.

[32] Sun Y, Shi Z, and Tamegai T 2019 *Supercond. Sci. Technol.* **32** 103001.

[33] Fang M H, Pham H M, Qian B, Liu T J, Vehstedt E K, Liu Y, Spinu L, and Mao Z Q 2008 *Phys. Rev. B* **78** 224503.

[34] Imai Y, Sawada Y, Nabeshima F, and Maeda A 2015 *Proc. Natl. Acad. Sci. USA* **112** 1937.

[35] Zhuang J, Yeoh W K, Cui X, Xu X, Du Y, Shi Z, Ringer S P, Wang X, and Dou S X 2014 *Sci. Rep.* **4** 7273.

[36] Terao K, Kashiwagi T, Shizu T, Klemm R A, and Kadowaki K 2019 *Phys. Rev. B* **100** 224516.

[37] Sun Y, Yamada T, Pyon S, and Tamegai T 2016 *Sci. Rep.* **6** 32290.

[38] Noji T, Suzuki T, Abe H, Adachi T, Kato M, and Koike Y 2010 *J. Phys. Soc. Jpn.* **79** 084711.

[39] Bean C P 1964 *Rev. Mod. Phys.* **36** 31-39.

[40] Sun Y, Pyon S, Tamegai T, Kobayashi R, Watashige T, Kasahara S, Matsuda Y, and Shibauchi T 2015 *Phys. Rev. B* **92** 144509.

[41] Sun Y, Taen T, Tsuchiya Y, Ding Q, Pyon S, Shi Z, and Tamegai T 2013 *Appl. Phys. Express* **6** 043101.

[42] Sun Y, Taen T, Tsuchiya Y, Pyon S, Shi Z, and Tamegai T 2013 *EPL (Europhysics Letters)* **103** 57013.

[43] Sun Y, Pyon S, Tamegai T, Kobayashi R, Watashige T, Kasahara S, Matsuda Y, Shibauchi T, and Kitamura H 2015 *Appl. Phys. Express* **8** 113102.

[44] Taen T, Nakajima Y, Tamegai T, and Kitamura H 2012 *Phys. Rev. B* **86** 094527.

[45] Taen T, Ohtake F, Pyon S, Tamegai T, and Kitamura H 2015 *Supercond. Sci. Technol.* **28** 085003.

[46] Lei H, Hu R, Choi E S, Warren J B, and Petrovic C 2010 *Phys. Rev. B* **81** 094518.

[47] Werthamer N R, Helfand E, and Hohenberg P C 1966 *Phys. Rev.* **147** 295.

[48] Xing X, Zhou W, Wang J, Zhu Z, Zhang Y, Zhou N, Qian B, Xu X, and Shi Z 2017 *Sci. Rep.* **7** 45943.

[49] Khim S, Lee B, Kim J W, Choi E S, Stewart G R, and Kim K H 2011 *Phys. Rev. B* **84** 104502.

[50] Maki K 1966 *Phys. Rev.* **148** 362-369.

[51] Kasahara S *et al* 2020 *Phys. Rev. Lett.* **124** 107001.

[52] Ok J M *et al* 2020 *Phys. Rev. B* **101** 224509.

[53] Nakayama K, Miyata Y, Phan G N, Sato T, Tanabe Y, Urata T, Tanigaki K, and Takahashi T 2014 *Phys. Rev. Lett.* **113** 237001.

[54] Yang H *et al* 2008 *Phys. Rev. Lett.* **101** 067001.

[55] Sun Y, Taen T, Yamada T, Pyon S, Nishizaki T, Shi Z, and Tamegai T 2014 *Phys. Rev. B* **89** 144512.

[56] Sun Y, Pyon S, and Tamegai T 2016 *Phys. Rev. B* **93** 104502.

[57] Rullier-Albenque F, Colson D, Forget A, and Alloul H 2012 *Phys. Rev. Lett.* **109** 187005.

[58] Lin H, Li Y, Deng Q, Xing J, Liu J, Zhu X, Yang H, and Wen H-H 2016 *Phys. Rev. B* **93** 144505.

[59] Chien T R, Wang Z Z, and Ong N P 1991 *Phys. Rev. Lett.* **67** 2088-2091.

[60] Tan S Y *et al* 2016 *Phys. Rev. B* **93** 104513.

[61] Abrikosov A A 1998 *Phys. Rev. B* **58** 2788.

[62] Smith R A Cambridge University Press, Cambridge, UK, 1978 Semiconductors.

[63] Kasahara S *et al* 2010 *Phys. Rev. B* **81** 184519.

[64] Watson M D *et al* 2015 *Phys. Rev. Lett.* **115** 027006.

[65] Nabeshima F, Ishikawa T, Shikama N, and Maeda A 2020 *Phys. Rev. B* **101** 184517.

[66] Luo N, and Miley G H 2002 *Physica C* **371** 259-269.

[67] Kohler M 1938 *Ann. Phys.* **424** 211.

[68] Eom M J, Na S W, Hoch C, Kremer R K, and Kim J S 2012 *Phys. Rev. B* **85** 024536.

[69] Harris J M, Yan Y F, Matl P, Ong N P, Anderson P W, Kimura T, and Kitazawa K 1995 *Phys. Rev. Lett.* **75** 1391.

[70] Li Q, Liu B T, Hu Y F, Chen J, Gao H, Shan L, Wen H H, Pogrebnyakov A V, Redwing J M, and Xi X X 2006 *Phys. Rev. Lett.* **96** 167003.

[71] Sun Y, Taen T, Yamada T, Tsuchiya Y, Pyon S, and Tamegai T 2015 *Supercond. Sci. Technol* **28** 044002.

[72] Xiang L, Kaluarachchi U S, Böhmer A E, Taufour V, Tanatar M A, Prozorov R, Bud'ko S L, and Canfield P C 2017 *Phys. Rev. B* **96** 024511.

[73] Holenstein S *et al* 2019 *Phys. Rev. Lett.* **123** 147001.

[74] Reiss P, Graf D, Haghighirad A A, Knafo W, Drigo L, Bristow M, Schofield A J, and Coldea A I 2019 *Nat. Phys.* **16** 89-94.

[75] Matsuura K *et al* 2017 *Nat. Commun.* **8** 1143.

[76] Chen G-Y, Wang E, Zhu X, and Wen H-H 2019 *Phys. Rev. B* **99** 054517.

[77] Gresty N C *et al* 2009 *J. Am. Chem. Soc* **131** 16944.

[78] Imai Y, Nabeshima F, and Maeda A 2017 *Condensed Matter* **2** 25.

[79] Mukasa K *et al* 2021 *Nat. Commun.* **12** 381.